\documentclass[acus]{JAC2003}

\usepackage{graphicx}
\usepackage{booktabs}
\usepackage{amsmath}

\begin{document}
\title{SIMULATIONS OF PION PRODUCTION IN THE DAEDALUS TARGET}

\author{Adriana Bungau, Roger Barlow, University of Huddersfield, Huddersfield, UK\\ 
Michael Shaevitz, Columbia University, New York, USA\\
Janet Conrad, Joshua Spitz, Tess Smidt, Massachusetts Institute of Technology, Massachusetts, USA}

\maketitle

\begin{abstract}
    DAE$\delta$ALUS, the Decay At-rest Experiment for $\delta_{CP}$ at a Laboratory for Underground Science will look for evidence of CP-violation in the neutrino sector, an ingredient in theories that seek to explain the matter/antimatter asymmetry in our universe. It will make a precision measurement of the oscillations of muon antineutrinos to electron antineutrinos using multiple neutrino sources created by low-cost compact cyclotrons. The experiment utilizes decay-at-rest neutrino beams produced by 800 MeV protons impinging a beam target of graphite and copper. Two well established Monte Carlo codes, MARS and GEANT4, have been used to optimise the design and the performance of the target. A study of the results obtained with these two codes is presented in this paper.
\end{abstract}

\section{INTRODUCTION}
Neutrino oscillation, the quantum mechanical process through which a neutrino created in one of three flavor states can be measured in another, is governed by three mixing angles, a CP-violating phase ($\delta_{CP}$), the neutrino mass differences, and the ratio of the experimental baseline to the neutrino energy ($L/E$). All of these parameters, except $\delta_{CP}$, have been measured. If $\delta_{CP}$ is non-zero, there is an asymmetry between how neutrinos and antineutrinos behave and this may help explain why there is more matter than antimatter in the universe.

DAE$\delta$ALUS~\cite{Alonso} will search for evidence of CP-violation in the neutrino sector by measuring the oscillation of muon antineutrinos into electron antineutrinos. DAE$\delta$ALUS will measure the oscillation probability at three distances and use these measurements to fit the probability formula and extract the $\delta_{CP}$ parameter. The experiment will use three neutrino sources and one large underground detector to make these measurements. Each neutrino source will consist of a compact H$_2^+$ cyclotron that will output 800 $\mathrm{MeV}$ protons into a beam target made of graphite and copper. While the graphite is the primary producer of pions in our target design, the copper effectively reduces backgrounds and transfers heat to keep the graphite below melting temperatures.

To design the DAE$\delta$ALUS targets, we simulate and compare the neutrino flux produced by various beam target geometries. Because the neutrino production rate, thus the pion production rate is our figure of merit, we must ensure that our simulations agree with experimental measurements of pion production. In this paper, we present a comparison of $\pi^+$ production in the simulation codes MARS and GEANT4 to available experimental data for protons with energies between 300 and 2500 $\mathrm{MeV}$ impinging on a graphite target. 

\section{TARGET DESIGN}

Proton collisions in the carbon target will create $\Delta$ resonances which then decay to create pions. The $\pi^+$ decay to create muon antineutrinos, the source of the oscillation signal, and the $\pi^-$ can decay to create electron antineutrinos, the dominant background for the experiment. Pions produced in the graphite travel into the copper prior to decaying. Once in the copper, the $\pi^-$s can be captured, preventing their decay. This reduces the ratio of electron antineutrino background to muon antineutrino signal to $\sim 10^{-3}$~\cite{burman}.

There are many interactions that occur in the beam target which ultimately affect the production of neutrinos. Protons lose energy as they traverse the target material prior to interacting, potentially preventing them from producing $\Delta$ resonances. Protons of sufficient energy can also re-interact causing multiple $\Delta$ resonances. Before considering these effects, it is important to benchmark the cross-sections that MARS and GEANT4 use for pion production in proton on carbon ($p\rightarrow C$) collisions.

\section{SIMULATION}

In experimental studies, $p\rightarrow C$ inelastic cross-sections and pion production cross-sections are measured using a thin carbon target, with thickness between 1 - 1.79 $\mathrm{g/cm^2}$ or 4.5 - 8 $\mathrm{mm}$~\cite{pion_0.73}~\cite{pion_0.585}~\cite{target1}. We simulate a 5 mm target for comparison to the experimental literature. While the DAE$\delta$ALUS targets will be on the order of 1 m thick, when benchmarking the two codes with a target thickness much smaller than the interaction length of the incident particles, we can neglect effects such as energy loss and re-interaction in the target.

In both the MARS and GEANT4 simulations, we count the total number of $\pi^{+}$ produced and number of inelastic proton interactions. We use these values and the simulation geometry and proton beam parameters to derive the cross-sections we compare to experimental data.

\subsection{MARS}

MARS~\cite{MARS} uses the Cascade Exciton Model (CEM) for materials with atomic mass A $>$ 3 and energy less than 5.0 $\mathrm{GeV}$. CEM considers nuclear reactions as proceeding through three states - cascade, pre-equilibrium, and equilibrium~\cite{CEM}. First, there is the intra-nuclear cascade where cascade particles are emitted leaving the residual nucleus in an excited state. Second, the nucleus relaxes according to the exciton model of the pre-equilibrium decay. Third, the equilibrium evaporative stage of the reaction occurs~\cite{CEM}~\cite{CEM_manual}.

\subsection{GEANT4}

Three independent models have been used in GEANT4~\cite{GEANT4} to simulate the inelastic proton interactions: the Bertini model, the Binary Cascade and the Li\`{e}ge (INCL) intra-nuclear cascade model coupled with the independent evaporation/fission code ABLA.

The Bertini Cascade Model~\cite{Heikkinen} generates the final state for hadron inelastic scattering by simulating the intra-nuclear cascade. Incident hadrons collide with protons and neutrons in the target nucleus and produce secondaries which in turn collide with other nucleons, the whole cascade being stopped when all the particles which can escape the nucleus have done so. Relativistic kinematics is applied throughout the cascade and the Pauli exclusion principle and conformity with the energy conservation law is checked. This model has been validated up to 10 GeV incident energy and is performing well for incident protons, neutrons, pions, photons and nuclear isotopes.

In the Binary Cascade Model~\cite{Binary} the propagation through the nucleus of the incident hadron and the secondaries it produces is modeled by a cascade series of two-particle collision, hence the name binary cascade. Between collisions the hadrons are transported in the field of the nucleus by Runge-Kutta method. The model is valid for incident protons, neutrons and pions and it reproduces detailed proton and neutron cross section data in the region 0-10 GeV.

To respond to the increasing user requirements from the nuclear physics community, the GEANT4 collaboration set a goal to complement the theory-driven models in this regime (the Bertini cascade and Binary cascade being the most widely used) with the inclusion of the INCL code also known as Liege cascade, often used with the evaporation/fission code ABLA~\cite{ABLA}. The code was validated recently against spallation data. It supports projectiles like protons in the energy range 200 MeV - 3 GeV. 

\section{RESULTS}

We provide two comparisons of simulation results to experimental data. First, we compare the $p\rightarrow C$ total inelastic cross-sections, $\sigma_{\mathrm{inelastic}}$, and the $\pi^+$ production total cross-sections, $\sigma_{\pi^+}$, for GEANT4, MARS and experiment. Second, we compare the ratio of these cross-sections. In this paper, we compare to the experimental pion production cross-sections from references~\cite{pion_0.73},~\cite{pion_0.585},~\cite{pion_1.6_2.5_3.5},~\cite{pion_0.400_0.450}, and~\cite{pion_0.330_0.400_0.500} and the experimental $p\rightarrow C$ inelastic scattering cross-section from references~\cite{proton_table} and~\cite{proton_table2}.

\begin{figure}[h]
    \centering
    \includegraphics*[width=80mm]{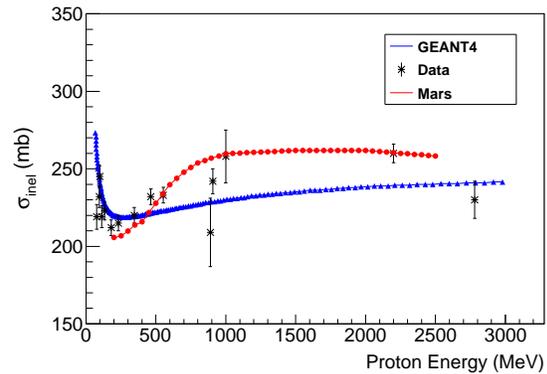}
    \caption{Plot of experimental and theoretical models' predictions for $\sigma_{\mathrm{inelastic}}$.}
    \label{inelastic}
\end{figure}

For the proton inelastic cross-section, the three theoretical models used in GEANT4 predict the same values. These values are compared with the MARS model and experimental data in Fig.~\ref{inelastic}. The GEANT4 models closely align with data for energies less than 1 $\mathrm{GeV}$. Above this energy, the data is highly varied, but is centered around both the GEANT4 and MARS models. Since these varying features do not reflect nature but rather experimental error, we use the GEANT4 models' prediction to represent the experimental data in the second half of our results.

\begin{figure}[h]
    \centering
    \includegraphics*[width=80mm]{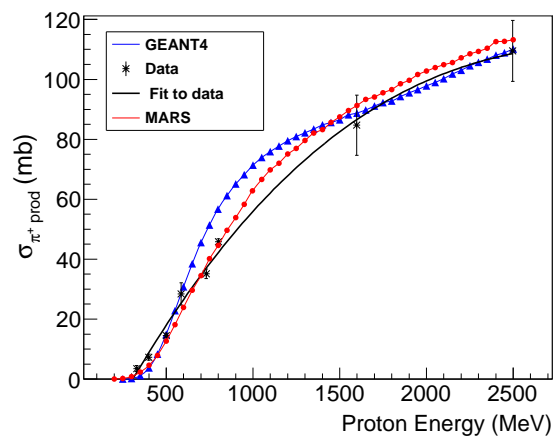}
    \caption{Plot of experimental and theoretical models predictions for $\sigma_{\pi^+}$.}
    \label{pionxsec}
\end{figure}

Out of the three GEANT4 models investigated, the Bertini model predictions for the total pion production cross-section are closest to the available experimental data. The comparison of the MARS and GEANT4 model predictions with experimental data is shown in Fig.~\ref{pionxsec}. Both codes compare well to the experimental values. We fit a third order polynomial to the experimental data to use in our final cross-section comparison.

The ratio of the number of protons producing $\pi^+$ and the number of protons which interact inelastically in our thin target is given by:  

\begin{equation}\label{eq:units}
    \frac{N_{\pi^{+}}}{N_p}={1-e^{-n\sigma_{\pi}x}\over 1-e^{-n\sigma_{p}x}}\approx \frac{\sigma_{\pi}}{\sigma_p}
\end{equation}

\noindent where $n$ is the number density of carbon atoms and $x$ is the thickness of the target. To obtain the righthand approximation, we Taylor expand around $x$ over $1/(n\sigma_{\pi})$ and $1/(n\sigma_{p})$, the interaction lengths of the two respective processes, to second order. We can make this approximation because the target thickness (5 $\mathrm{mm}$) is much smaller than the interaction length of either process ($\sim$1 m).

\begin{figure}[h]
    \centering
    \includegraphics*[width=80mm]{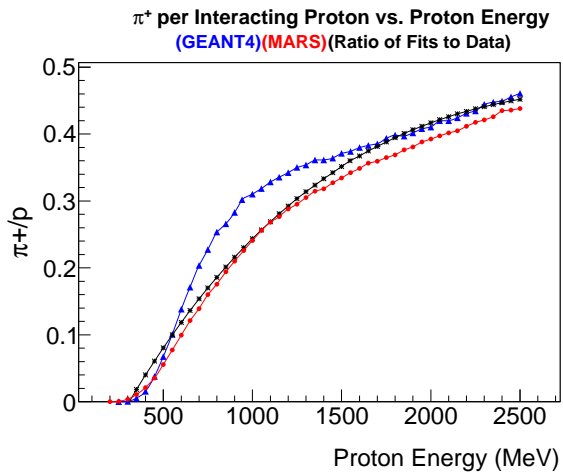}
    \caption{Plot of MARS and GEANT simulation results and the ratio given by the two fits through the experimental data.
}
    \label{interp}
\end{figure}

In Fig.~\ref{interp}, the GEANT4 and MARS cross-section ratios are obtained by dividing the simulation values in Fig.~\ref{pionxsec} by the simulation values in Fig.~\ref{inelastic}. The ``Ratio of Fits to Data'' is obtained by dividing the third order polynomial fit to the experimental data in Fig.~\ref{pionxsec} by the GEANT4 model prediction in Fig.~\ref{inelastic}. 

\section{CONCLUSION}

Pion production calibration is crucial for ensuring accurate simulation of the neutrino fluxes created by the DAE$\delta$ALUS sources. Calibrating for energies lower than the initial proton energy is important for simulating the pion production in thicker targets where more protons interact but do so at lower than their initial energies. 

From this study we find that MARS and GEANT4, for the models presented, consistently agree on pion production rates within 30\% and that both codes compare well to experimental data in the energy region of interest, $<$~800~$\mathrm{MeV}$. The information from this study can be used to reweight simulation data, such that simulated flux more accurately reflects experimental data.

The neutrino flux predictions we make with these simulations will be used to decide on experimental parameters such as cyclotron power cycle and determine the required running-time needed to achieve DAE$\delta$ALUS's physics goals.

\end{document}